\documentclass[twoside,11pt,onehalfspacing]{article}

\usepackage{fullpage}
\usepackage{authblk}
\usepackage{natbib}
\usepackage{graphicx}
\pdfoutput=1

\usepackage{booktabs}
\usepackage{subfigure}
\usepackage{tabularx}
\usepackage{multirow}
\usepackage{xcolor}
\usepackage{tikz}
\usepackage{amsmath}

\usepackage{lipsum}
\usepackage{caption}
\usepackage{subfigure}
\usepackage{amsfonts}
\usepackage{ulem}
\usepackage{xcolor}
\usepackage{multirow}
\usepackage{graphicx}
\usepackage{amsmath}
\usepackage{amssymb}
\usepackage[small,compact]{titlesec}
\usepackage{amsmath}
\usepackage{siunitx} 
\usepackage{booktabs}
\usepackage{float}
\usepackage{accents}
\usepackage{hhline}
\usepackage{float}
\restylefloat{table}
\floatstyle{plaintop}
\restylefloat{table}
\usepackage{tabularx}
\usepackage[section]{placeins}
\def\ab#1{{\bf \color{blue}#1}}
\def\ad#1{{\bf \color{orange}#1}}

\def\rafid#1{{\bf \color{blue}RM: #1}}

\def\ab#1{#1}
\def\ad#1{#1}
\def\rafid#1{#1}

\begin{document}

\title{The importance of evaluating the complete automated knowledge-based planning pipeline}

  \renewcommand*{\Authands}{, }
    \author[1]{Aaron Babier}
  \author[1]{ Rafid Mahmood}
  \author[2]{Andrea L. McNiven}
  \author[3]{Adam Diamant}
  \author[1]{Timothy C. Y. Chan\vspace*{-0.25cm}}
  \affil[1]{Department of Mechanical \& Industrial Engineering, University of Toronto}
  \affil[2]{Radiation Medicine Program, Princess Margaret Cancer Centre}
  \affil[3]{Schulich School of Business, York University\vspace*{-0.5cm}}

\maketitle

\begin{abstract}
We determine how prediction methods combine with optimization methods in two-stage knowledge-based planning (KBP) pipelines to produce radiation therapy treatment plans. We trained two dose prediction methods, a generative adversarial network (GAN) and a random forest (RF) with the same 130 treatment plans. The models were applied to 87 out-of-sample patients to create two sets of predicted dose distributions that were used as input to two optimization models. The first optimization model, inverse planning (IP), estimates weights for dose-objectives from a predicted dose distribution and generates new plans using conventional inverse planning. The second optimization model, dose mimicking (DM), minimizes the sum of one-sided quadratic penalties between the predictions and the generated plans using several dose-objectives. Altogether, four KBP pipelines (GAN-IP, GAN-DM, RF-IP, and RF-DM) were constructed and benchmarked against the corresponding clinical plans using clinical criteria; the error of both prediction methods was also evaluated. The best performing plans were GAN-IP plans, which satisfied the same criteria as their corresponding clinical plans (78\%) more often than any other KBP pipeline. However, GAN did not necessarily provide the best prediction for the second-stage optimization models. Specifically, both the RF-IP and RF-DM plans satisfied all clinical criteria 25\%  and 15\% more often than GAN-DM plans (the worst performing planning), respectively. GAN predictions also had a higher mean absolute error (3.9 Gy) than those from RF (3.6 Gy). We find that state-of-the-art prediction methods when paired with different optimization algorithms, produce treatment plans with considerable variation in quality.

\end{abstract}

\section{Introduction}
\begin{figure}[t]
\centering
         \includegraphics[width=0.93\linewidth]{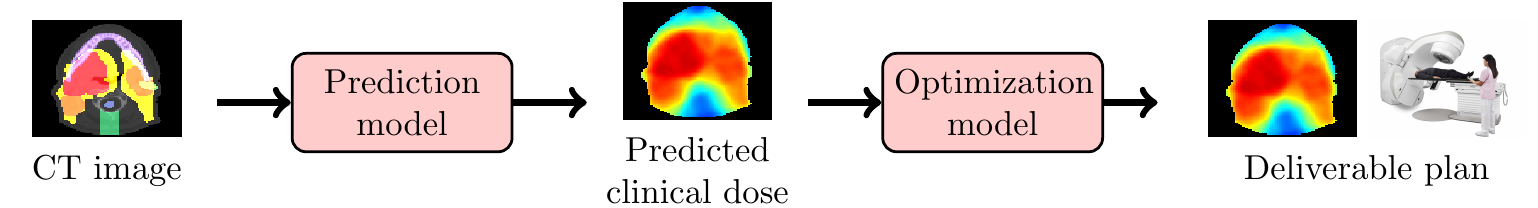}
        \caption{Overview of the automated knowledge-based  planning pipeline.}
        \label{fig:pipeline}
\end{figure}

\ad{Automated knowledge-based planning (KBP) is a data-driven approach that uses previous radiation therapy treatments to generate high quality plans for patients diagnosed with cancer. KBP is typically conceptualized as a two-stage pipeline (see Figure~\ref{fig:pipeline}). In the first stage, a machine learning (ML) model uses contoured CT images to predict the dose that should be delivered to a patient. In the second stage, an optimization model uses the dose prediction from the first stage to generate fluence maps \ab{or a set of beam
apertures}.} In the past decade, there has been significant research in improving KBP, focusing either on advancing the machine learning or the optimization stage \ad{independently of each other} \citep{Wu:2009aa,PCAZhu:2011aa, PCAYuan:2012aa,appenzoller:2012predicting, Yang:2013aa,Shiraishi:2015aa,Shiraishi:2016ab,Kearney:2018aa,Fan:2018aa,Nguyen:2019aa}. 
\ad{In this work, we explore whether the interaction between the prediction and optimization model affects the overall quality of the final plans.}

Several prediction models have been developed that can accurately \ad{predict aspects of a clinical dose distribution from a patient's anatomy.} Originally, these prediction models required the engineering of useful features from patient information and only predicted simple summaries (e.g., desirable DVHs) that could be used to design objectives for inverse planning software \citep{Wu:2009aa,PCAZhu:2011aa, PCAYuan:2012aa,appenzoller:2012predicting, Yang:2013aa,Shiraishi:2015aa,Shiraishi:2016ab,Babier:2018b}. However, modern deep learning techniques can learn useful features in order to predict dose directly from CT images \citep{GANCER,Kearney:2018aa,Fan:2018aa,Nguyen:2019aa,Babier:2019}. These high-dimensional predictions contain more information and permit better integration into more sophisticated automated KBP pipelines \citep{GANCER}.

The dominant optimization \ad{models} for KBP are inverse planning (IP) \citep{Babier:2018a} and dose mimicking (DM) \citep{petersson:2016}. \rafid{The choice of the parameters, objectives, and constraints in these models can lead to final treatment plans with characteristics that differ significantly from the initial predictions. For example, a prediction model may produce dose distributions that consistently predict excess dose to an OAR, but an optimization model with an objective to minimize the dose to that OAR may be able to correct for this bias. As a result, \ad{prediction models that produce dose distributions with good criteria satisfaction may not necessarily produce final plans with the same properties.} Constructing effective automated KBP pipelines, therefore, requires careful selection of both the prediction and optimization model.} 

\begin{figure}[t]
\centering
         \includegraphics[width=0.8\linewidth]{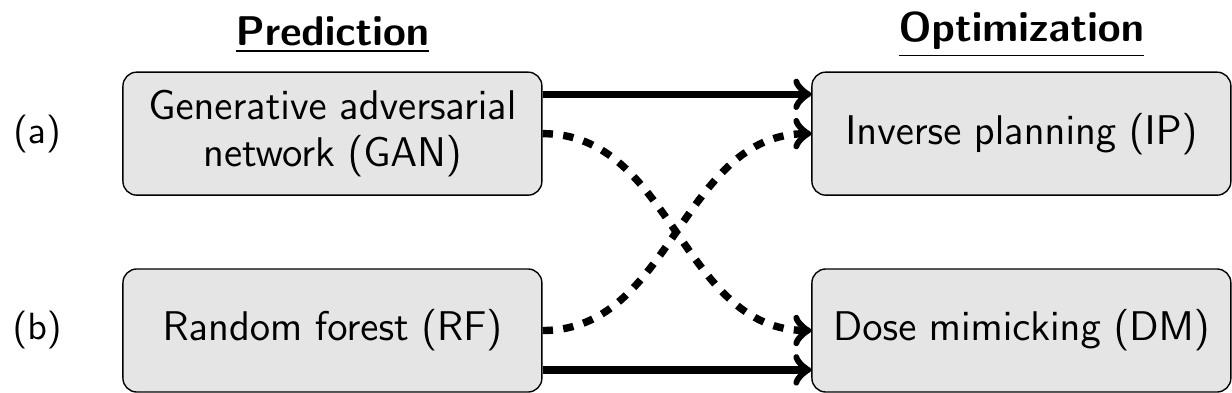}
        \caption{Overview of the automated knowledge-based planning pipelines evaluated in this paper. Solid lines connect the prediction and optimization methods that have been tested together in (a) \cite{Babier:2019}  and (b)  \cite{McIntosh:2017aa}; dashed lines connect the methods that have not been tested in the extant literature.}
        \label{fig:kbp_combos}
\end{figure}

In this paper, we perform the first comparison between different combinations of prediction and optimization models in KBP; \ab{each model was previously validated in a KBP pipeline} \citep{McIntosh:2017ab,Babier:2019}. In total, we consider two dose prediction methods---a generative adversarial network \citep{Babier:2019} and a random forest \citep{McIntosh:2017aa}---and two optimization methods---inverse planning \citep{Babier:2018a} and dose mimicking \citep{petersson:2016}. We then evaluate the four corresponding KBP pipelines (see Figure~\ref{fig:kbp_combos}) using a large dataset of 217 patients with oropharyngeal cancer. \ab{We observe that the choice of} \ad{both the prediction and optimization model can significantly affect the quality of the final plans generated by a KBP pipeline.} 




\section{Methods and Material}
\ad{We used CT images with contours, which highlight the regions-of-interest (ROIs), and dose distributions from clinically accepted treatment plans to train two dose prediction models that were then tested on out-of-sample clinical plans. The resulting predicted dose distributions were then passed through each optimization model to generate fluence-based treatment plans. Figure~\ref{fig:kbp_combos} gives an overview of the pipelines, which were then evaluated in terms of the quality of plans they produced.}


\subsection{Data}
For this research ethics board approved study, we obtained plans for 217 oropharyngeal cancer treatments 
delivered at a single institution with 6 MV, step-and-shoot, intensity-modulated radiation therapy. All plans were prescribed 70 Gy and 56 Gy in 35 fractions to the gross disease (PTV70) and elective target volumes (PTV56), respectively; in 130 plans there was also a prescription of 63 Gy to the intermediate-risk target volume (PTV63). The organs-at-risk (OARs) were the brainstem, spinal cord, right parotid, left parotid, larynx, esophagus, mandible, and the limPostNeck, which is an artificial structure used to limit dose to the posterior neck. 

\subsection{Prediction models}
We trained two state-of-the-art dose prediction models with the same 130 plans from our dataset and used the remaining 87 for out-of-sample testing. 

\subsubsection{Generative adversarial network}
\begin{figure}[t]
\centering
        \includegraphics[width=1\linewidth]{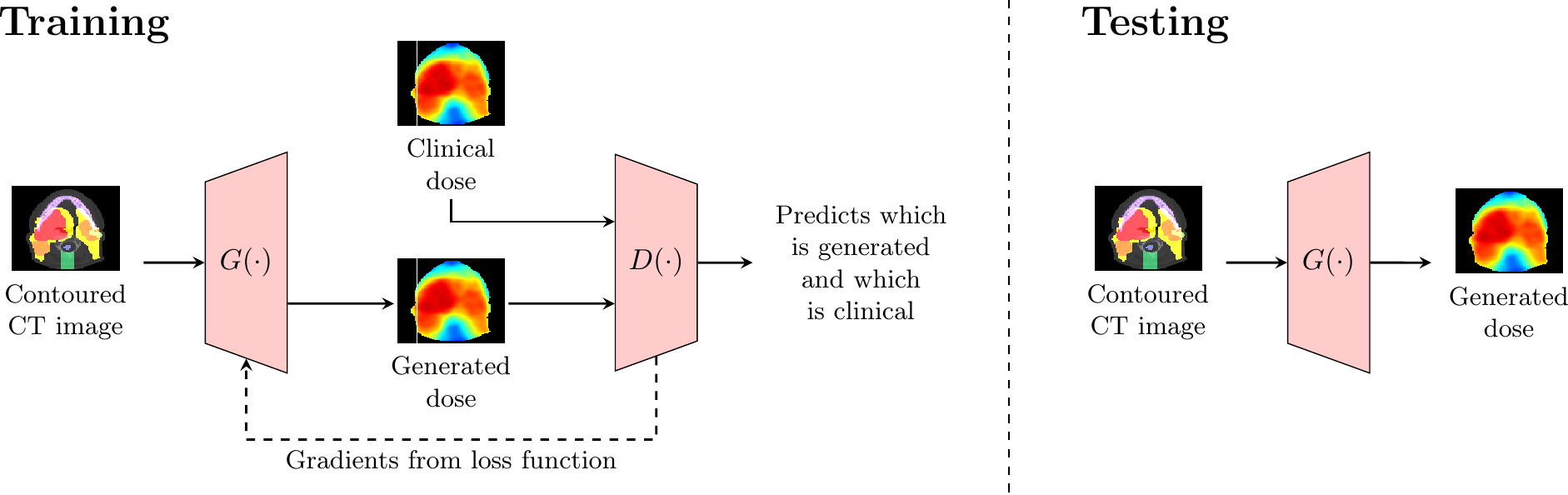} 
        \caption{Overview of GAN training and testing phases.}
          \label{fig:gan_training}
\end{figure}

\ad{Our conditional generative adversarial network (GAN) model \citep{isola:2017image} is based on \cite{Babier:2019} and uses two convolutional neural networks: (1) a generator that produces a dose distribution from a contoured CT image; and (2) a discriminator that tries to differentiate between the artificially generated dose and the actual clinical dose (see Figure~\ref{fig:gan_training}). The generator is trained to minimize the mean absolute difference between the artificially generated image and the ground truth (i.e., clinical dose). The objective is regularized by the discriminator to make the output of the generator indistinguishable from a real clinical dose distribution. We then normalize the resulting dose generated by GAN so that it satisfies all target criteria.}

\subsubsection{Random forest}

The random forest model is a slight variation of the RF from \cite{McIntosh:2017ab}. It uses the 148 features summarized in Table~\ref{rf_features} 
to predict the dose delivered to each voxel independently. Of these features, 122 were generated by applying Gaussian filters (GFs) to the grayscale CT images. One of the GFs was isotropic ($\sigma = 10$) and a second was the Laplacian of the Gaussian ($\sigma = 10$). The remaining 120 filters were made from all combinations of the following four parameters: (a) first and second order GFs; (b) $\sigma = 4, 12, 24, 48,$ and $64$; (c) rotations of $0, 90, 180,$ and $270$ degrees; and (d) rotations in each of the three axes. \ad{RF was trained to minimize the mean squared difference between the prediction and the ground truth using the default settings of  \textsf{randomForestRegressor} from \texttt{scikit-learn}}.

\begin{table}[t]
\centering
	\begin{tabularx}{\linewidth}{llX} \toprule 
	 Feature & \ab{Quantity} & Description \\ \midrule
     Structure & 11 & Structure voxel is classified as (one-hot-encoded)\\
     $x$-coordinate & 1 & Voxel's positions on the $x$-axis in a slice\\
     $y$-coordinate & 1 & Voxel's positions on the $y$-axis in a slice\\
     $z$-coordinate & 1 & Plane of voxel's slice\\
     ROI distance & 11 & Voxel's distance from surface of each ROI \\
     CT gray-scale & 1 & Voxel's gray-scale in the CT image\\
     GF CT gray-scale & 122 & Voxel's gray-scale in CT image post GF\\ \bottomrule
	\end{tabularx}
\caption{The features used in RF to predict the dose for each voxel.}
\label{rf_features}
\end{table}

\subsection{Optimization models}
\ad{\ab{For all out-of-sample patients}, the output from each prediction model was passed to both optimization models which were solved using \texttt{Gurobi 7.5.}} The complexity of all generated treatment plans was constrained to a sum-of-positive-gradients (SPG) value of 55~\citep{Craft:2007aa}. SPG was used since it is a convex surrogate for the physical deliverability of a plan and the parameter 55 was chosen as it is two standard deviations above the average clinical SPG~\citep{Babier:2018b}. Both optimization models used the same set of 
targets $\mathcal{T}$ and healthy structures $\mathcal{I}$. \ad{Each target $t \in \mathcal{T}$ was a planning target volume (PTV) with a prescribed dose $\theta^t$. The healthy structures contained in $\mathcal{I}$ were the brainstem, spinal cord, right parotid, left parotid, larynx, esophagus, mandible, and limPostNeck. Each target structure $t \in \mathcal{T}$ and healthy structure $i \in \mathcal{I}$ was divided into a set of voxels $\mathcal{O}^t$ and $\mathcal{O}^i$, respectively.}

The KBP-generated plans were delivered from nine equidistant coplanar beams at angles 0$^\circ$, 40$^\circ$, \ldots, 320$^\circ$. Those beams were divided into a set of beamlets $\mathcal{B}$, which make up one fluence map at each beam angle. The relationship between the intensity $w_b$ of beamlet $b$ and dose $d_v$ deposited to voxel $v$ was determined using the influence matrix $D_{v,b}$ generated by the \textsf{IMRTP} library from \texttt{A\ Computational\ Environment\ for\ Radiotherapy\ Research}~\citep{CERR}, and it is given by
\begin{equation*}
\label{IM}
d_v = \sum\limits_{b \in \mathcal{B}} D_{v,b}w_b.
\end{equation*}
\subsubsection{Inverse planning}

We followed a  previously developed two-stage approach to inverse planning~\citep{Babier:2018a}. In the first step, we estimate the objective weights for a conventional inverse planning model that makes a predicted dose distribution optimal. In the second step, the estimated weights are used to re-solve the conventional inverse planning optimization model and construct a treatment plan.
The objective to be minimized was a sum of 65 functions: seven per OAR and three per target. The objectives for the OARs were the mean dose, maximum dose, and the average dose above 0.25, 0.50, 0.75, 0.90, and 0.975 of the maximum predicted dose to the OAR. The objectives for the target were the maximum dose, average dose below prescription, and average dose above prescription.

\subsubsection{Dose mimicking}
Our dose mimicking (DM) model minimized the sum of one-sided penalties to generate a plan that performs as close as possible to the predicted dose on several voxel- and structure-based objectives. \ad{Two types of OAR objectives were used}. The first was a voxel-based objective that minimizes the dose $d_v$ that exceeds the predicted dose $\hat{d}_v$ for each voxel $v$:
\begin{equation}
x_v = \text{max}\left\{0,\ d_v - \hat{d}_v\right\}, \quad \forall v \in \mathcal{O}^i, \forall i \in \mathcal{I}.\label{dm:oar_mean}
\end{equation}
The second was a structure-based objective that minimizes the maximum dose until it no longer exceeds the maximum predicted dose:
\begin{equation}
y^i = \text{max}\left\{0,\  \underset{v\in\mathcal{O}^i}{\text{max}}\left\{d_v\right\} - \underset{v\in\mathcal{O}^i}{\text{max}}\{\hat{d}_v\}\right\}, \quad \forall i \in \mathcal{I}.\label{dm:oar_max}
\end{equation}

\ad{Three types of target objectives were also used}. The first was a voxel-based objective to minimize the average deviation \emph{below} the prescribed target dose $\theta^t$, which is the average \emph{underdose} to target $t$. Specifically, the objective function $l_v$ penalizes dose until the plan underdose is no worse than what was predicted for each voxel $v$. The objective is formulated as:
\begin{equation}
l_v = \text{max}\left\{0,\ \theta^{t} - d_v - \text{max}\{0,\  \theta^{t} - d_v\}\right\}, \quad \forall v \in \mathcal{O}^t, \forall t \in \mathcal{T}.\label{dm:underdose}\\
\end{equation}
\ad{Similarly, the second objective was also voxel-based}, however, it minimizes the average deviation \emph{above} the prescribed target dose $\theta^t$, which is the average \emph{overdose} to target $t$. Specifically, the objective function $u_v$ penalizes dose until the plan overdose is no worse than what was predicted for each voxel $v$. The objective is formulated as: 
\begin{equation}
u_v = \text{max}\left\{0,\ d_v - \theta^{t} - \text{max}\{0,\  \hat{d}_v - \theta^{t}\}\right\}, \quad \forall v \in \mathcal{O}^t, \forall t \in \mathcal{T}.\label{dm:overdose}\\
\end{equation}
\ad{The final target objective was structure-based. It maximizes the minimum dose to the target until it exceeds the minimum dose that was predicted for the target:}
\begin{equation}
z^t = \text{min}\left\{0,\  \underset{v\in\mathcal{O}^i}{\text{min}}\{\hat{d}_v\} - \underset{v\in\mathcal{O}^i}{\text{min}}\left\{d_v\right\}\right\}, \quad \forall t \in \mathcal{T}.\label{dm:max_tar}
\end{equation}
To form the dose mimicking optimization problem, we constructed linearized forms of equations \eqref{dm:oar_mean}-\eqref{dm:max_tar} by introducing appropriate auxiliary variables and constraints and summed those terms in the objective function. We divided each voxel-based objective by the number of voxels in its respective structure. The conceptual DM model is
%
\begin{equation*}
\begin{aligned}
& \underset{x, y, l, u, z, w}{\text{minimize}}
& & \sum\limits_{i\in \mathcal{I}}\left(\frac{1}{|\mathcal{O}^i|}\sum\limits_{v\in \mathcal{O}^i}x_v^2 + (y^i)^2\right) 
+ \sum\limits_{t\in\mathcal{T}}\left(\frac{1}{|\mathcal{O}^t|}\sum\limits_{v\in \mathcal{O}^t}\left(l_v^2 + u_v^2\right) + (z^t)^2\right),\\
& \text{subject to}
& & \eqref{dm:oar_mean} - \eqref{dm:max_tar},\\
& & & SPG \le 55.
\end{aligned}
\end{equation*}

\subsection{Performance analysis}

We evaluated four distinct KBP pipelines based on the plans they produced; predictions were also evaluated because they are an important intermediate step. We refer to the four sets of KBP plans as GAN-IP, RF-IP, GAN-DM, and RF-DM. \ad{We evaluated the predicted and plan dose distributions in terms clinical criteria, the difference in the performance of each optimization model when the same set of predictions is used as input, and the prediction error. Details of these performance metrics are presented below.}


\subsubsection*{Criteria Satisfaction}

We quantified the quality of KBP plans by how often they satisfied the same clinical criteria presented in Table~\ref{goals}. Specifically, we examined how often the plans satisfied the same criteria as the clinical plans in each class of criteria (i.e., OARs, targets, and all ROIs, which includes both OARs and targets). \ad{Finally, we evaluated the quality of the prediction models to determine whether criteria satisfaction in the predicted dose distribution} is an early indicator of final plan quality.


\begin{table}[t]
\caption{The planning criteria used for evaluation: $\mathcal{D}_{99}$ is the minimum dose to 99\% of the structure volume, $\mathcal{D}_{mean}$ is the mean dose to a structure, and $\mathcal{D}_{max}$ is the maximum dose to a structure.}
\centering
	\begin{tabular}{l  c}
	 \toprule
	 Structure & Criteria \\ \midrule
	Brainstem & $\mathcal{D}_{max} \le$ 54 Gy \\
	Spinal Cord & $\mathcal{D}_{max} \le$ 48 Gy \\
   	Right Parotid & $\mathcal{D}_{mean}\le$ 26 Gy \\
	Left Parotid & $\mathcal{D}_{mean}\le$ 26 Gy \\
 	Larynx & $\mathcal{D}_{mean} \le$ 45 Gy \\
  	Esophagus & $\mathcal{D}_{mean} \le$ 45 Gy \\
	Mandible & $\mathcal{D}_{max}\le$ 73.5 Gy \\
	PTV56 & $\mathcal{D}_{99}\ \ge$ 53.2 Gy \\  
  	PTV63 & $\mathcal{D}_{99}\ \ge$ 59.9 Gy \\
	 PTV70 & $\mathcal{D}_{99}\ \ge$ 66.5 Gy \\\bottomrule
	\end{tabular}
\label{goals}
\end{table}

\subsubsection*{Optimization performance differences}
For each clinical planning criterion (Table~\ref{goals}), \ad{we evaluated the difference in dose between plans generated with an identical set of predictions but a different optimization model; the differences} between the two optimization models (i.e., IP and DM) are visualized with a box plot. We then used a two-sided Mann-Whitney U test to determine if plans generated by IP were the same (null hypothesis) or different (alternative hypothesis) from those generated by DM for the population of plans generated from each set of predictions. For these and all subsequent hypothesis tests, $p < 0.01$ was considered significant.

\subsubsection*{Prediction performance differences}
We evaluated the error of each prediction method by evaluating the median absolute difference between the predicted and clinical dose distributions across each ROI for every out-of-sample plan. \ad{The error is visualized with a box plot and we used a two-sided Mann-Whitney U test to determine if  GAN had the same (null hypothesis) or a different (alternative hypothesis) prediction error than RF.} 

\section{Results}
\subsubsection*{Criteria Satisfaction}
Table~\ref{clinCrit} summarizes the performance of the predicted and plan dose distributions. RF-DM plans achieved similar OAR criteria satisfaction to the clinical plans most often (83.9\%). However, GAN-IP plans satisfied the target criteria 28.8\% more often than RF-DM plans, and achieved close to RF-DM performance on the OAR criteria (4.6 percentage points less). Across all ROIs, the proportion of GAN-IP plans satisfied the same criteria as the corresponding clinical plans 17\% more often than its closest competitor (RF-IP).  Additionally, while GAN-IP performed better than RF-IP, GAN-DM performed worse than RF-DM, which suggests that \ad{there is an interaction effect between the prediction and optimization model that must be accounted for.}


In Table~\ref{clinCrit}, we also compare the predictions to the clinical plans. We \ad{emphasize that unlike the generated plans, i.e., IP and DM plans, the predictions} are only an intermediate step in the KBP pipeline. Here, we found that GAN predictions \ad{exhibited poor performance on all OAR criteria} (24.1\%) which we attribute to the poor performance on the mandible criteria (13.6\%). The performance of RF and GAN predictions over all target criteria was similar. Overall, RF 
predicted that plans could satisfy the same criteria as the clinical plans in 78.2\% of cases, which \ad{far exceeded} GAN predictions (24.1\%). \ad{Most importantly, however, these results do not carry through to the final plans. That is, only GAN-IP plans achieved the same proportion of All ROI criteria (78.2\%) that was predicted by RF.}

\begin{table}[t] 			
\caption{The percentage of plans that satisfied the same clinical criteria as the clinical plans. \ad{Only IP and DM plans use the full KBP pipeline}}									
\begin{tabular}{c l | c c | c c |  c c } 			\toprule											
&		&	\multicolumn{2}{c|}{Predictions}			&	\multicolumn{2}{c|}{IP plans}			&	\multicolumn{2}{c}{DM plans}			 \\
&		&	GAN	&	RF	&	GAN-IP	&	RF-IP	&	GAN-DM	&	RF-DM	 \\ \midrule
\parbox[t]{4mm}{\multirow{7}{*}{\rotatebox[origin=c]{90}{OARs}}} 														
&	Brainstem		&	100.0	&	98.9		&	100.0	&	100.0	&	100.0	&	98.9	 \\
&	Spinal Cord	&	97.7		&	100.0	&	100.0	&	100.0	&	90.8		&	97.7	 \\
&	Right Parotid	&	64.7		&	64.7		&	94.1		&	76.5		&	76.5		&	82.4	 \\
&	Left Parotid	&	81.8		&	54.5		&	81.8		&	54.5		&	81.8		&	90.9	 \\
&	Larynx		&	71.4		&	89.8		&	91.8		&	87.8		&	81.6		&	93.9	 \\
&	Esophagus	&	100.0	&	100.0	&	100.0	&	100.0	&	100.0	&	100.0	 \\
&	Mandible		&	13.6		&	100.0	&	81.8		&	74.2		&	36.4		&	93.9	 \\ \midrule
\parbox[t]{4mm}{\multirow{3}{*}{\rotatebox[origin=c]{90}{Targets}}} 														
&	PTV56		&	100.0	&	97.8		&	97.8		&	95.7		&	100.0	&	93.5	 \\ 
&	PTV63		&	100.0	&	98.0		&	100.0	&	98.0		&	100.0	&	100.0	 \\
&	PTV70		&	100.0	&	100.0	&	100.0	&	98.3		&	98.3		&	58.6	 \\ \midrule\midrule
\parbox[t]{4mm}{\multirow{3}{*}{\rotatebox[origin=c]{90}{All}}} 														
&	OARs		&	24.1		&	80.5		&	79.3		&	69.0		&	41.4		&	83.9	 \\
&	Targets		&	100.0	&	97.7		&	98.9		&	95.4		&	98.9		&	70.1 	 \\
&	\textbf{ROIs}			&	\textbf{24.1}		&	\textbf{78.2}		&	\textbf{78.2}		&	\textbf{66.7}		&	\textbf{41.4}		&	\textbf{56.3}	 \\ \bottomrule
\label{clinCrit} 				
\end{tabular} 	
\end{table} 														

\subsubsection*{Optimization performance differences}

\begin{figure}[t]
\centering
\subfigure[\ GAN plan differences]{
\includegraphics[width=0.47\linewidth]{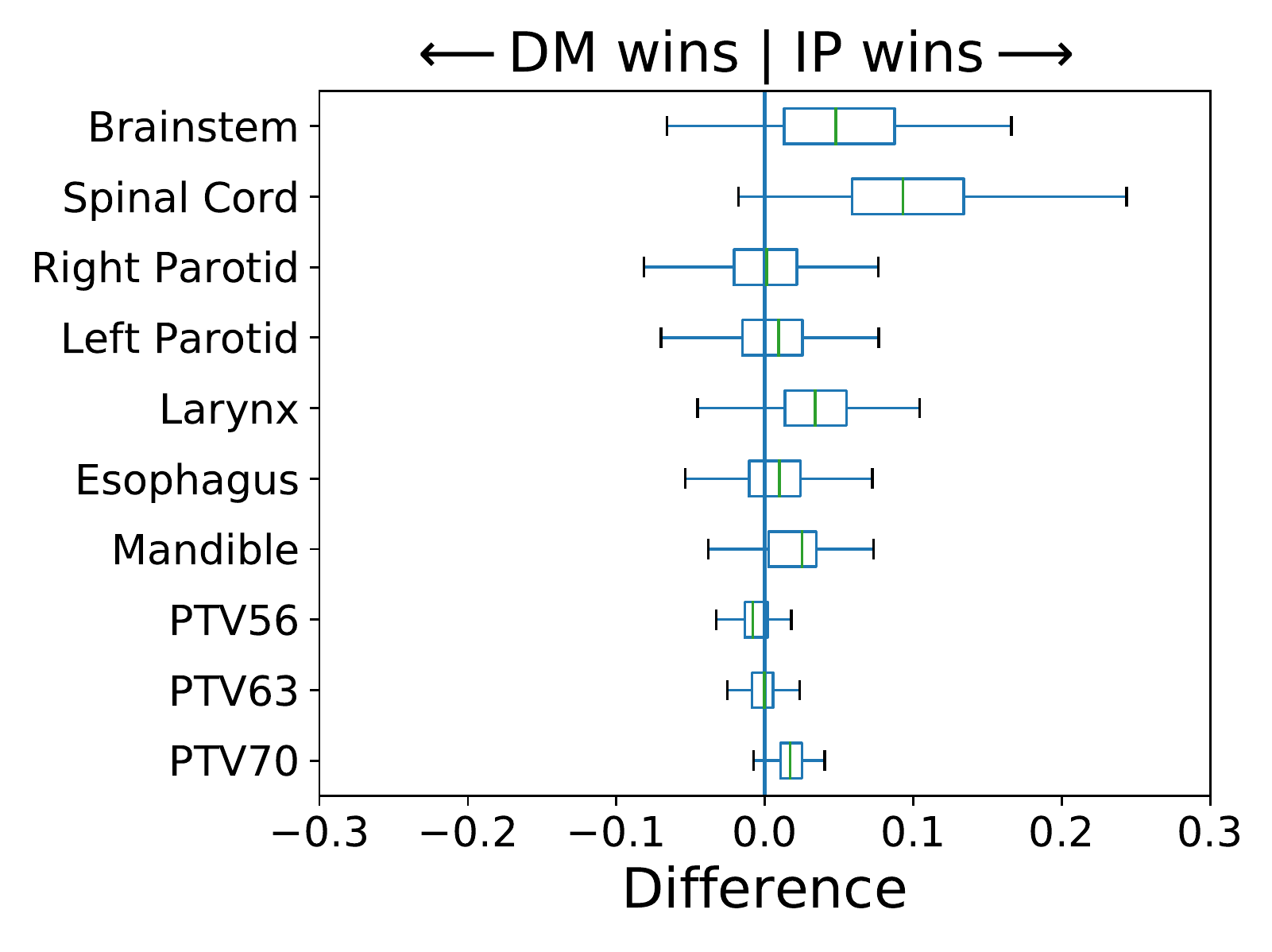} }
\subfigure[\ RF plan differences]{
\includegraphics[width=0.47\linewidth]{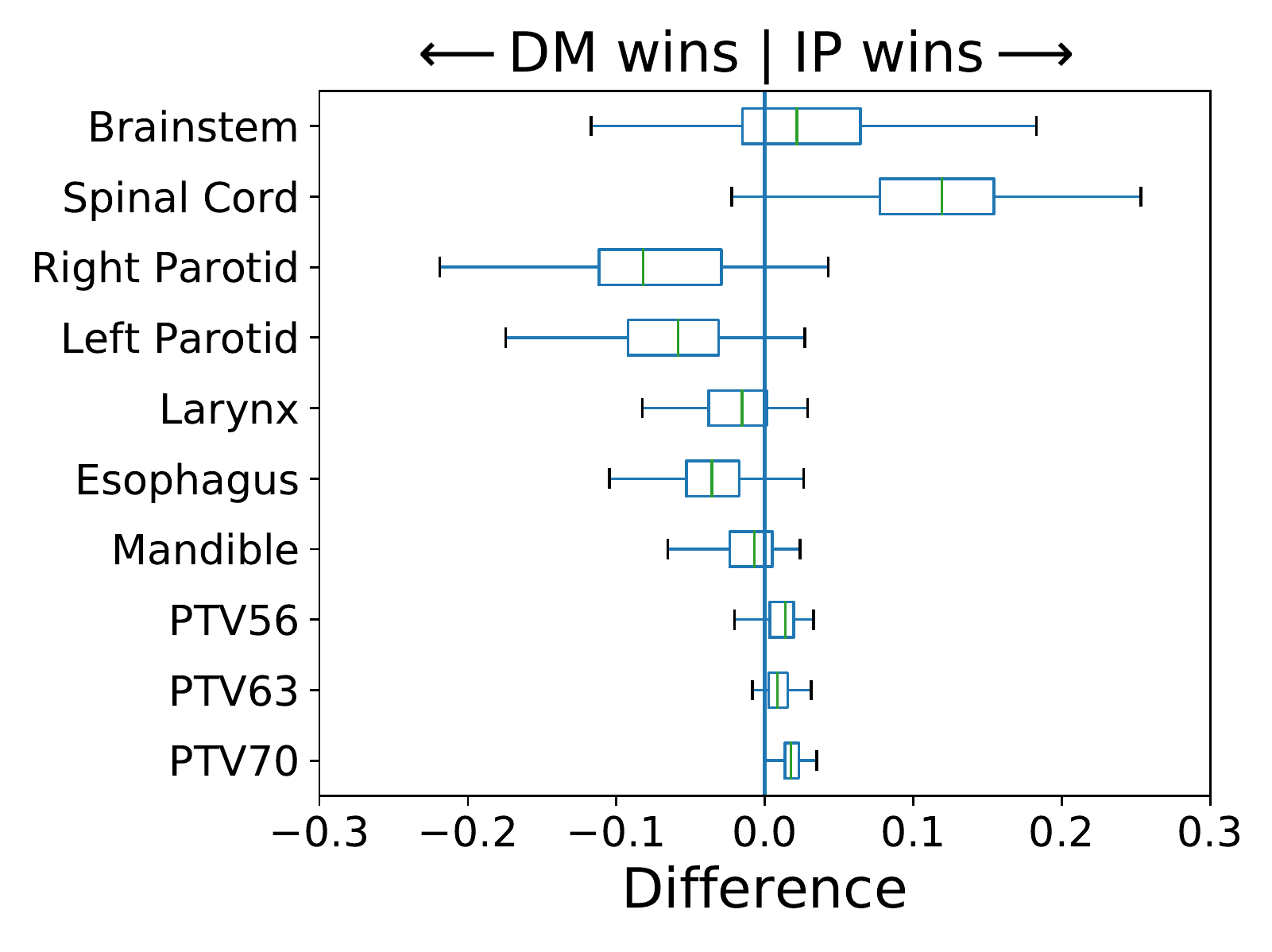} }
\caption{The difference in terms of clinical planning criteria between plans generated by IP and DM where the input to both models are (a) GAN predictions and (b) RF predictions. Positive differences imply that the IP plan was better than the DM plan in that criterion. The boxes indicate median and interquartile range (IQR). Whiskers extend to the minimum of 1.5 times the IQR and the most extreme outlier.}
\label{figure:opt_results}
\end{figure}

In Figure~\ref{figure:opt_results}, we present a box plot to compare the quality of plans from \ad{different optimization models when the same prediction model was used as input.} The plot shows how the plans generated by IP differ from those generated by DM in terms of the \ad{dose delivered to each clinical planning criterion relative to the dose threshold of that criterion}. On average, IP was better than DM by 2\% when GAN predictions were used as input. \ad{However, we found no difference between plans generated by IP and DM when RF predictions were used as input.}
\ad{We also found that IP performed better than DM in 69.5\% and 50.8\% of all evaluation criteria \ab{when the inputs were from GAN and RF}, respectively.} Statistically, when the GAN predictions were used as input, the plans generated by IP and DM \ad{performed differently on the clinical criteria ($p<0.001$). However, we observed no difference ($p=0.045$) when RF predictions were used as input to the optimization models. Overall, we observed that the performance of each optimization model was dependent on the prediction model that was used. } 

\subsubsection*{Prediction performance differences}
In Figure~\ref{figure:prediction_error}, we present the distribution of mean absolute differences between the predicted and clinical dose over the regions of interest (i.e., the mean absolute error between the predictions and clinical plans). Although both models had the same median prediction error across all OARs (4.3 Gy), RF error across targets (1.3 Gy) \ad{was much} lower than GAN \ad{error} (3.0 Gy). Overall, GAN predictions had higher median error across all ROIs (3.9 Gy) than RF predictions (3.6 Gy), and these predictions errors were significantly different ($p<0.001$).

\definecolor{babyBlue}{HTML}{1f77b4}

\begin{figure}[t]
\centering
\includegraphics[width=0.7\linewidth]{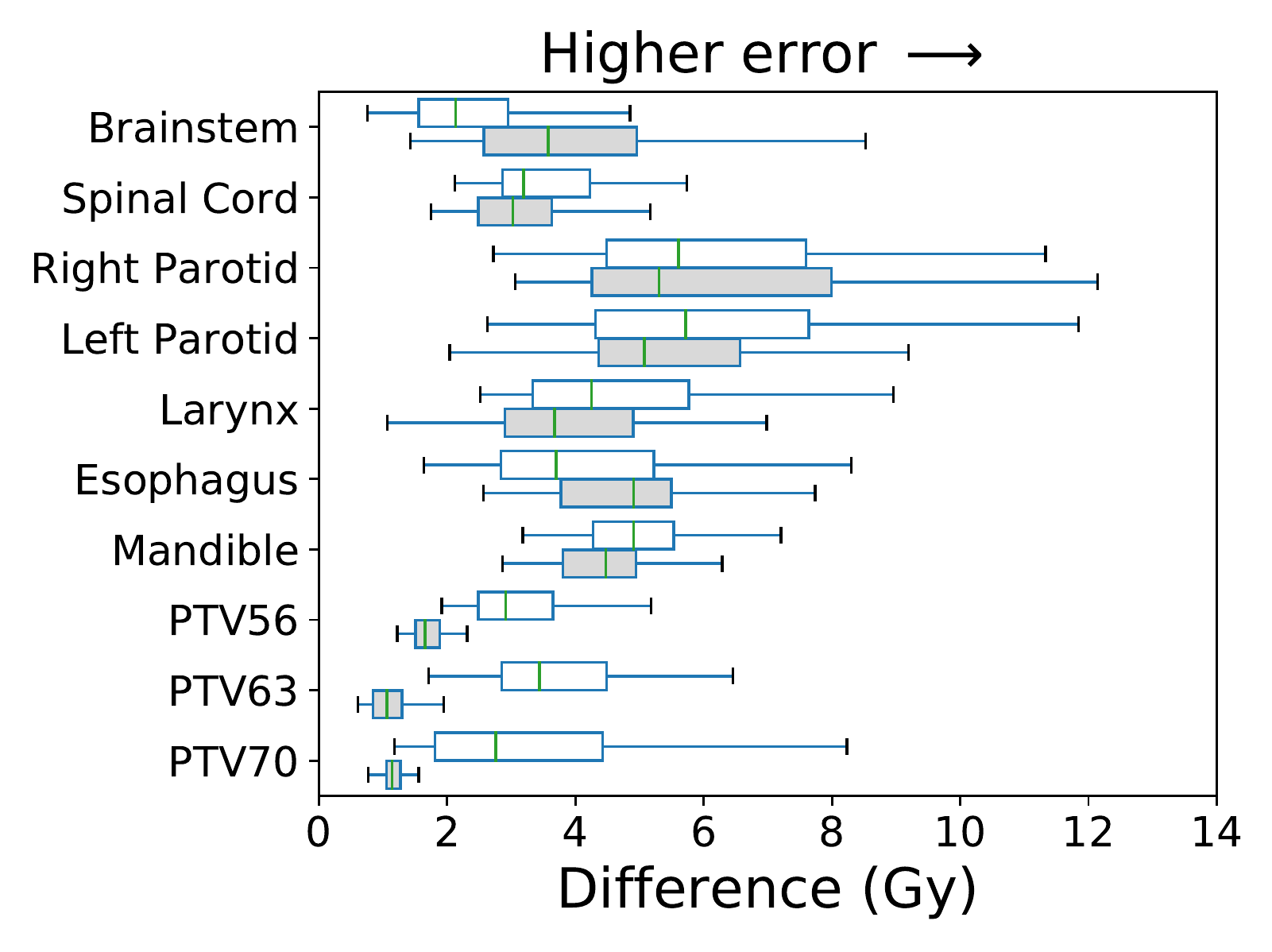}\\
GAN predictions \begin{tikzpicture} \draw[fill=white!25,line width=0.5pt,draw=babyBlue] rectangle (0.4,0.2);\end{tikzpicture}
RF predictions \begin{tikzpicture}\draw[fill=black!25,line width=0.5pt,draw=babyBlue]   rectangle (0.4,0.2); \end{tikzpicture}\quad
\caption{The distribution of average dose differences between KBP prediction and clinical dose over all ROIs. The boxes indicate median and IQR. Whiskers extend to the minimum of 1.5 times the IQR and the most extreme outlier.}
\label{figure:prediction_error}
\end{figure}
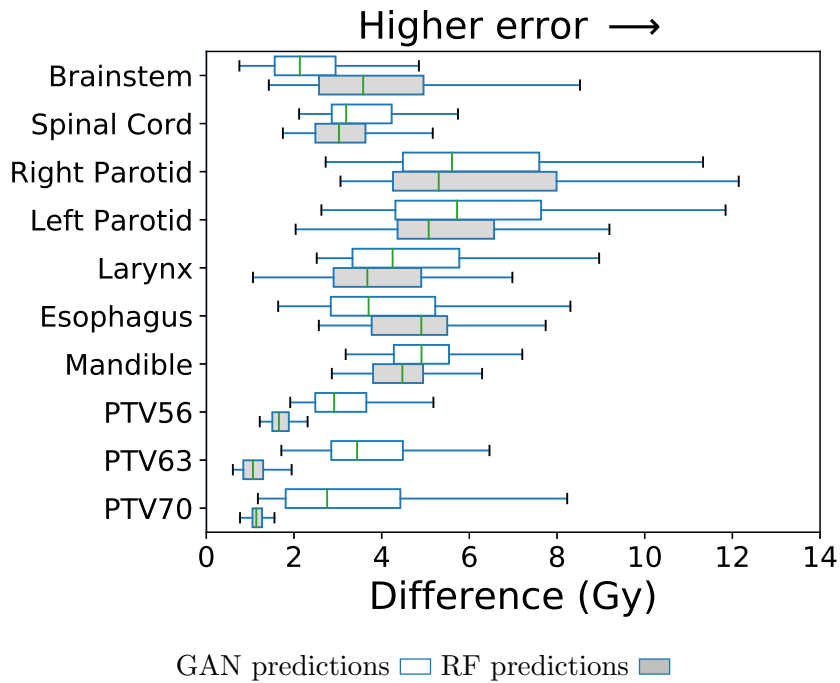

\section{Discussion}
Historically, \ad{each stage of KBP has} been developed in isolation with a focus on improving the prediction stage. In this paper, we show that there are interaction effects between the prediction and optimization stages of KBP that significantly affect the quality of the generated plans. \ad{Our experimental} setup consists of four KBP pipelines that were assembled from \ad{two existing KBP methods, i.e., \cite{Babier:2019}  and \cite{McIntosh:2017aa} (see Figure~\ref{fig:kbp_combos})}. \ab{Overall, the best performing combination of prediction and optimization methods was the GAN and IP.} However, we also demonstrate that \ad{predictions that produce good plans with one optimization model (e.g., GAN-IP) do not always produce good plans with another optimization model} (e.g., GAN-DM). 

Although both RF and GAN predict 3D dose distributions, they differ in their approach. RF predicts the dose to each voxel 
independently of every other voxel. In contrast, GAN predicts the dose to all voxels simultaneously, thereby making predictions that are conditioned on the predictions of neighboring voxels. RF generally produces predictions that are more similar to clinical plans on summary statistics like mean absolute dose difference (Figure~\ref{figure:prediction_error}). This is likely because GAN optimizes a regularized loss function that encourages realistic looking images. \ad{This results in \ab{predictions} that have} worse performance on summary statistics as compared to a model like RF that minimizes the squared difference between predictions and the ground truth without regularization. 

The quality of deliverable plans depends heavily on the combination of the prediction and optimization components used to construct the KBP pipeline. For example, combining GAN and IP \ad{results in} plans that perform well on average in terms of satisfying clinical criteria. Interestingly, the KBP pipelines that perform the best contain stages that use same order loss and objective functions (i.e., linear-linear or quadratic-quadratic).
Namely, GAN (trained with mean absolute loss) and IP (optimized with a linear objective function) produce the best IP plans. Similarly, RF (trained with mean squared loss) and DM (optimized with a quadratic objective function) produce the best DM plans.

When considering OAR and target criteria satisfaction as the two key components, we observe that there is no single two-stage KBP pipeline that \ad{dominates all others}. While GAN-IP performs at least as well as RF-IP and GAN-DM on both metrics, RF-DM outperforms GAN-IP on OAR criteria satisfaction; we conjecture that this is because GAN predictions perform poorly on the mandible criterion. 
Inverse planning includes a specific objective that minimizes the dose to the mandible, so even if the predictions (incorrectly) assume that mandible criterion satisfaction is unimportant, the mandible objective in IP improves mandible criterion performance. In contrast, dose mimicking attempts to construct dose distributions that are no worse than the predictions (in terms of Equations \eqref{dm:oar_mean} - \eqref{dm:max_tar}), which generally leads to less improvement on the mandible criterion. Due to the biased nature of GAN predictions towards the mandible, IP can help to improve the single weak criterion with minimal expense to the criteria. On the other hand, IP is unable to exploit any significant under-performance in RF predictions, which generally perform well across most criteria.

\ad{A limitation of our work is that, although we identified that the prediction and optimization stages of KBP affect the overall quality of the plans they generate, we were unable to isolate the root cause of those effects. A second limitation is that the computational resources required for this analysis scales exponentially with the number of prediction and optimization models considered. As a result, it is computationally intensive to determine what existing optimization model should be paired with a new approach to dose prediction (and vice versa).}



\section{Conclusion}
This study \ad{demonstrates} that the performance of an automated KBP pipeline is dependent on how well the prediction and optimization models perform together. As a result, we recommend that new prediction methods should be tested with multiple optimization models before they are considered to be state-of-the-art (and vice versa).

\section{Acknowledgments}
This research was supported in part by the Natural Sciences and Engineering Research Council of Canada.

\bibliography{refsForArXiv}

\end{document}